# Cooperative Ru(4*d*)-Ho(4*f*) magnetic orderings and phase coexistence in the 6H-perovskite multiferroic Ba$_3$HoRu$_2$O$_9$

T. Basu[1,2], V. Caignaert[2], F. Damay[3], T.W. Heitmann[4], B. Raveau[2], X. Ke[1]

[1]*Department of Physics and Astronomy, Michigan State University, East Lansing, MI 48824, USA*

[2]*Normandie Univ, ENSICAEN, UNICAEN, CNRS, CRISMAT, 14000 Caen, France*

[3]*Laboratoire Léon Brillouin, Grenoble, France*

[4]*Missouri Research Reactor, University of Missouri, Columbia, Missouri 65211, USA*

**We report cooperative magnetic orderings in a 6H-perovskite multiferroic system, Ba$_3$HoRu$_2$O$_9$, via comprehensive neutron powder diffraction measurements. This system undergoes long-range antiferromagnetic ordering at $T_{N1}$ ~ 50 K with a propagation wave vector of K$_1$ = (0.5 0 0), a transition temperature much higher than the previously reported one at ~10 K ($T_{N2}$). Both Ru and Ho-moments order simultaneously below $T_{N1}$, followed by spin-reorientations at lower temperatures, demonstrating strong Ru(4*d*)-Ho(4*f*) magnetic correlation. Below $T_{N1}$ another magnetic phase with a propagation wave vector K$_2$ = (0.25 0.25 0) emerges and coexists with the one associated with K$_1$, which is rarely observed and suggests complex magnetism due to phase competition in the magnetic ground state. We argue that the exchange-striction arising from the up-up-down-down spin structure associated with K$_2$-wave vector below $T_{N2}$ may be responsible for the small ferroelectric polarization reported previously in this compound.**

The electron and magnetic correlation of *d* and *f* electrons has been a core research topic in condensed matter physics, and it plays a decisive role in determining materials properties, such as unconventional superconductivity, metal-insulator transition, magnetoresistance, multiferroicity, as well as a rich variety of magnetic orderings. Particularly, the strong *d-f* magnetic correlations with competing magnetic interactions often give further interesting properties. For instance, the compounds containing magnetic rare-earth (*R*)-ions and transition-metal (TM) ions, e.g. *R*MnO$_3$,[1] *R*Mn$_2$O$_5$,[1–3] *R*$_2$BaNiO$_5$ [4,5] undergo simultaneous ordering of TM and *R*-moments due to 3*d-4f* magnetic correlation, exhibiting intriguing multiferroicity / strong magnetoelectric coupling.

While there have been extensive studies on the materials exhibiting 3*d*-4*f* coupling in recent years, much fewer reports in the literature exist on materials composed of both heavy R and 4*d* TM ions that could potentially possess strong 4*d*-4*f* coupling,[6,7] specifically considering the fact that 4*d*/5*d* electron orbitals are of special interest due to their compelling effects of large spin-orbit coupling and extended *d*-orbitals. The fascinating pyrochlore ruthenates *R*$_2$Ru$_2$O$_7$ exhibit Ru$^{4+}$ ordering at high temperature followed by the ordering of rare-earth ions at lower temperature induced by the 4*d-4f* coupling.[8–10] However, unlike other *R*- members, the magnetic ordering of Er and Ru in Er$_2$Ru$_2$O$_7$ is still ambiguous, where effects of strong magnetic anisotropy and crystal-field effect are speculated.[11,12] The double perovskites $A_2R$RuO$_6$ (A=Sr, Ba) exhibit successive magnetic ordering of Ru$^{5+}$ at high temperature followed by $R$-ordering at lower temperature for $R$=Ho, Er, whereas, simultaneous magnetic ordering is observed for $R$=Nd and Dy.[13–15] On other hand, only one magnetic anomaly is reported for Ba$_4R$Ru$_3$O$_{12}$ system where the role of Ru$^{4+}$/Ru$^{5+}$ and $R^{3+}$ions remains unclear.[16] Therefore, the $d$-$f$ magnetic correlation is always intriguing for different $R$-ions not only for compounds in the same family but also for systems with distinct crystal structures / space groups. Another system of current interest is 6H-perovskite Ba$_3$BB′$_2$O$_9$ (B= 3d transition metal/ Sr/Ca/Na/ Lanthanides, B′= 4$d$/5$d$ metal like Ru, Nb, Sb, Ir), which exhibits versatile exotic properties depending on the nature of B and B′ ions, such as, dimer system, geometrical frustration, quantum spin-liquid, charge-ordering, unusual valence state, multiferroicity, etc.[17–26]

The system Ba$_3R$Ru$_2$O$_9$, crystallizing in 6H-perovskite structure, consists of Ru$_2$O$_9$ dimers (face-sharing RuO$_6$-octahedra) which are interconnected by corner-sharing MO$_6$ octahedra and possess an average valence state of 4.5 of Ru-ion when $R = R^{3+}$. Recently, we have reported magnetodielectric (MD) coupling for Ba$_3R$Ru$_2$O$_9$, which is significantly enhanced for heavier rare-earth member Ba$_3$HoRu$_2$O$_9$.[25,27] Such an enhanced MD coupling and the emergent ferroelectricity in Ba$_3$HoRu$_2$O$_9$ were speculated to arise from stronger 4$d$-4$f$ magnetic correlation between Ru and heavy $R$-ions.[25] The light rare-earth compound Ba$_3$NdRu$_2$O$_9$ exhibits a ferromagnetic ordering of Nd-moments at $T_c$ ~ 24 K, followed by an antiferromagnetic (AFM) ordering of Ru$_2$O$_9$ dimer ~18 K and canted AFM ordering of Nd ~17 K.[20,28] In contrast, the heavy-rare-earth members ($R$=Tb, Gd, Ho, Er, etc) undergo AFM ordering at low temperature ~10 K, which is ascribed to the ordering of rare-earth ions, without any further magnetic ordering down to 2 K.[29]. Until now, there is no report of a detailed magnetic structure of this Ba$_3R$Ru$_2$O$_9$ system for any heavy-rare-earth member in this series, which is warranted in order to confirm the speculation of strong 4$d$-4$f$ magnetic correlation for heavy $R$-members.

In this Rapid Communication, via comprehensive neutron powder diffraction measurements, we report simultaneous magnetic ordering of Ru$^{4+}$/Ru$^{5+}$ and Ho$^{+3}$ moments in Ba$_3$HoRu$_2$O$_9$ at $T_{N1}$ ~ 50 K, which is ascribed to strong 4$d$-4$f$ magnetic correlation. A rare phase-co-existence of two different magnetic structure with **K$_1$** = (0.5 0 0) and **K$_2$** = (0.25 0.25 0) is revealed below $T_{N2}$ ~ 10 K, arising from competing exchange interactions. The up-up-down-down spin structure associated with K$_2$ is likely intimated with the ferroelectricity below $T_{N2}$ of this compound.

High quality Ba$_3$HoRu$_2$O$_9$ polycrystalline samples were synthesized using solid state chemistry method as described in our earlier report.[25] Magnetic susceptibility measurements were conducted using Superconducting Quantum Interference Device (SQUID) magnetometer, and heat capacity measurements were done using Physical Properties Measurements System (PPMS), both produced from Quantum Design. Neutron powder diffraction measurements were carried out using a two-axis-diffractometer G4.1 with an incident neutron wavelength of 2.425 Å in LLB (Laboratoire Léon Brillouin), France and a triple-axis spectrometer (TRIAX) with incident neutron wavelength of 2.359Å at the University of Missouri Research Reactor. The magnetic structure was resolved using Fullprof and SARAh program.[30,31]

The inset of Fig.1(a) shows the temperature dependence of dc magnetic susceptibility $\chi$ measured in the presence of 1T magnetic field. The drop in $\chi$ below 10 K (assigned as $T_{N2}$) indicates an antiferromagnetic (AFM) phase transition, which agrees well with the previous reports.[25,29] The inverse susceptibility (Fig. S1 in Supplemental Material (SM))[32] deviates from linearity (Curie-Weiss behavior) below ~ 100 K, implying the presence of (short-range) magnetic correlation in this system far above $T_{N2}$. Figure 1(a) presents the specific heat divided by temperature ($C/T$) in the presence of $H = 0$ and 5 T dc magnetic field as a function of temperature. The $C/T$ is nearly constant down to 50 K from high temperature (also see Fig. S1 in SM for a broader view)[32], then slowly decreases with lowering the temperature till around 13 K, followed by a $\lambda$-shape anomaly around 10 K (Fig. 1a) in the absence of magnetic field, which confirms the long-range magnetic ordering at $T_{N2}$. In the presence of 5 T magnetic field, the feature at $T_{N2}$ shifts to lower temperature (~ 8 K), consistent with the AFM nature of this system. Interestingly, one can see that the curve measured at $H = 0$ and 5 T starts to bifurcate below ~ 45 K, further suggesting the presence of magnetic-correlation at much higher temperature compared to $T_{N2}$.

In order to have a better understanding of the magnetic ordering of $Ba_3HoRu_2O_9$, we have performed neutron power diffraction measurements. Figure 1(b) shows the diffraction intensity as a function of momentum transfer Q measured at several temperatures ranging from 1.5 to 80 K. The insets present an expanded view at Q = 0.93 and 1.07 Å$^{-1}$. There are several important features worth pointing out. i) There is no change of nuclear Bragg peaks (also see Fig. S2 in SM)[32] at all temperatures measured, which indicates no structural phase transition down to 1.5 K. ii) Below $T_{N2}$, for instance at $T$ = 8.1 K and 1.5 K, there are extra peaks showing up at Q = 1.07, 0.93, 0.75 and 0.69 Å$^{-1}$ compared to the data measured at $T$ = 80 K, indicating their magnetic nature and as to be discussed next, the wave vectors associated with these Q values are (0.5 0 2), (0.75 -0.25 1), (0.5 0 1), and (0.25 0.25 1) respectively, iii) Intriguingly, as shown in the inset of Fig. 1(b), the magnetic Bragg peak at Q = 1.07 Å$^{-1}$ persists even above $T_{N2}$, for instance, at $T$ = 11.3 K, 19 K, and 30 K, while magnetic Bragg peaks at other Q values disappear. This suggests the presence of another magnetic ordering existing at between 30 K to 80 K, which is far above the previously reported magnetic transition at ~ 10 K ($T_{N2}$). To obtain magnetic ordering temperatures, Figure 1(c) and its inset presents the temperature dependence of scattering intensities measured at aforementioned four Q values. In contrast to other magnetic Bragg peaks whose intensity drops to the background signal at $T_{N2}$, the Bragg peak intensity of (0.5 0 2) peak sharply decreases when increasing temperature from 2 K(Fig.1c), become nearly constant around $T_{N2}$, followed by a gradual drop above 15 K until it reaches the background signal around 50 K ($T_{N1}$). These features clearly indicate that the system undergoes two magnetic phase transitions, one at $T_{N1}$ ~ 50 K and the other at $T_{N2}$ ~10 K.

Rietveld refinement of the neutron diffraction data measured at several temperatures using SARAh and Fullprof program are presented in Fig. 2. The nuclear scattering data at 80 K (Fig. 2a) is well fitted with the space group P6$_3$/mmc, which affirms the high crystalline quality of the sample. We do not find any impurity via Rietveld refinement within the resolution limit of the instrument (<2%). For magnetic refinement, the possible propagation (**K**)-vectors associated with this space group are listed in Table-S1 in SM[32]. We identify the magnetic propagation vector of the neutron diffraction measured at 30 K and 11.3 K (i.e., $T_{N1} < T < T_{N2}$) to be **K$_1$** = (0.5 0 0). Considering the space group P6$_3$/mmc and **K$_1$** = (0.5 0 0), there are 4-irreducible representations associated with Ho atom, represented by $\Gamma_{mag}$ (Ho)= $1\Gamma_1^1 + 0\Gamma_2^1 + 2\Gamma_3^1 + 0\Gamma_4^1 + 1\Gamma_5^1 + 0\Gamma_6^1 + 2\Gamma_7^1 + 0\Gamma_8^1$ and 8-irreducible representation associated with Ru atom, represented by $\Gamma_{mag}$ (Ru)= $1\Gamma_1^1 + 2\Gamma_2^1 + 2\Gamma_3^1 + 1\Gamma_4^1 + 1\Gamma_5^1 + 2\Gamma_6^1 + 2\Gamma_7^1 + 1\Gamma_8^1$ (see Table S2 and S3 in SM[32]). As both Ru and Ho-atoms have ordered magnetic moments, their irreducible representations should contain any of the combinations of $\Gamma_1$, $\Gamma_3$, $\Gamma_5$ and $\Gamma_7$. Among these four representations, $\Gamma_7$-model gives the best fit (Fig. 2b and 2c). We found that the magnetic peak profile is best modeled only when both Ho and Ru-atoms have non-zero magnetic moment (detailed are discussed in SM[32], see Fig.S2 in SM[32]). Below $T_{N2}$ we find that the propagation of the magnetic Bragg peaks can be indexed with a propagation vectors of **K$_1$** = (0.5 0 0) and **K$_2$** = (0.25 0.25 0), indicating that two different magnetic phases coexist. Fig. 2d shows the neutron scattering data measured at $T$ = 8.1 K and the refinement results. Considering the space group P6$_3$/mmc and **K$_2$** = (0.25 0.25 0), there are 4-irreducible representations associated with Ho atom represented by $\Gamma_{mag}$ (Ho) = $1\Gamma_1^1 + 1\Gamma_2^1 + 2\Gamma_3^1 + 2\Gamma_4^1$, and there are 4-irreducible representations associated with Ru atom, represented by $\Gamma_{mag}$ (Ru) = $3\Gamma_1^1 + 3\Gamma_2^1 + 3\Gamma_3^1 + 3\Gamma_4^1$ (see Table S4 and S5 in SM[32]). We find that a combination of $\Gamma_1$ (for **K$_2$** = (0.25 0.25 0)) and $\Gamma_7$ (for **K$_1$** = (0.5 0 0)) models gives the best refinement result below $T_{N2}$. The Rietveld refinement for 1.5 K data are shown in Figure S4 in SM[32]. This infers that two different magnetic phases coexist below $T_{N2}$. The neutron peak-shape is modeled with a Thompson-Cox-Hastings pseudo-Voigt function in Fullprof program which is convolution of Lorentzian and Gaussian functions. The peak shape of the magnetic Bragg reflections associated with the **K$_1$** vector is Gaussian (negligible Lorentzian part), whereas, the magnetic Bragg reflections associated with the **K$_2$** vector has a small Lorentzian component along with a Gaussian one, which implies that magnetic correlation length associated with the **K$_2$** structure is shorter than that associated with the **K$_1$** structure.

The magnetic structures associated with these two magnetic phases are depicted in Fig. 3. At $T_{N2} < T < T_{N1}$, for instance, $T$ = 30 K, both Ho and Ru spins are ordered ferromagnetically along the b-axis and antiferromagnetically in the ac-plane (AFM along the a-axis and canted AFM along the c-axis), as shown in Fig. 3(a). The Ho spins are nearly aligned in the ab-plane with a slight tilting towards the c-axis, whereas the Ru spins are nearly aligned along the c-axis with a tilting towards the ab-plane. With lowering the temperature, the magnetic structure remains nearly the same but with more tilting of both Ho and Ru moments towards the c-axis and an enhanced moment size (Fig.3b for 11.3 K and Table-1). This is indicative of strong 4*d*-4*f* magnetic correlation in this system. The slow rotation of Ru and Ho-moments is observed with lowering the temperature. However, the tilting of Ho-moments toward c-axis is enhanced negligibly at 11K compared to that of 19 K, whereas the change in canting angle observed between 19 and 30 K is larger (see Table-1). This results in a gradual but anomalous change in the intensity of (0.5 0 2) magnetic peak (Fig. 1c) due to the change in magnetic structure factor in addition to the enhancement of the magnetic moment value. Below $T_{N2}$, there is a change in magnetic structure. Figures 3 (c,d) represent the magnetic structures associated with **K₁** = (0.5 0 0) and **K₂** = (0.25 0.25 0) propagation wave vectors respectively by refining the neutron scattering data measured at $T$ = 8.1 K. As illustrated in Fig. 3(c), the magnetic structure associated with **K₁** remains nearly the same below and above $T_{N2}$. Nevertheless, the component of Ho magnetic moment ($M_c$) along the c-axis is significantly enhanced below $T_{N2}$, compared to the $M_a$ and $M_b$ components along a and b axes (see Table-1 in SM[32]). In contrast, $M_c$ of the Ru-magnetic moment is significantly reduced, compared to the $M_a$ and $M_b$ components, which is distinct from the scenario observed above $T_{N2}$. Interestingly, such spin reorientation of both Ho and Ru moments gives rise to the extinction of the (0.5 0 1) and (0.5 -1 1) magnetic Bragg peaks above $T_{N2}$ while the (0.5 0 2) magnetic Bragg peak persists. The calculated and experimentally obtained intensity of the magnetic peaks are listed in Table-S6 in SM[32]. Figure 3 (d) shows the refined magnetic structure associated with **K₂**. For this magnetic phase, both Ho and Ru spins are completely aligned in the ab-plane with an up-up-down-down antiferromagnetic structure, while spins are ferromagnetically aligned along the c-axis. Note that the total magnetic moment of Ho-atom is nearly saturated (~10.1 $\mu_B$) at $T$ = 8.1 K, and the total magnetic moment of Ru is about 1.6 $\mu_B$. The magnetic structure remains same down to $T$ = 1.5 K (Figure S5 and S6 in SM[32]). The smooth increase of the magnetic Brag peak intensity below $T_{N2}$ (main panel and inset of Fig.1c) is due to the fact of the enhancement of magnetic moments toward saturation. The application of high magnetic field may flip the spin along $H$ and may transform the up-up-down-down spin-structure to the up-up-up-down spin-structure. The $H$-induced magnetic transition with weak hysteresis around ~3 T in isothermal magnetization (Ref.[25]) could be attributable to a spin-flip transition.

The observation of two magnetic orderings in $Ba_3HoRu_2O_9$ is quite intriguing. First, no clear anomaly at the onset of long-range ordering (LRO) around 50 K ($T_{N1}$) is observed in either bulk magnetic susceptibility or heat capacity measurements, though, the Gaussian nature of the peak-shape of magnetic reflections associated with **K₁** confirms the LRO below $T_{N1}$. This is presumably because the magnetic ordering at $T_{N1}$ is weak in nature, where both Ho and Ru spins start to order but with small magnetic moment. Because of short-range magnetic correlation above $T_{N1}$, entropy starts to vary slowly from high temperature, crosses over with a minimal change around $T_{N1}$ due to weak magnetic ordering, followed by further gradual changes due to continuous slow spin-saturation and spin-reorientation with further lowering temperature. Despite the absence of anomaly in heat capacity at $T_{N1}$, one can clearly observe the onset of a bifurcation of heat capacity measured at zero and high magnetic field as shown in Fig. 1(a). Right below $T_{N2}$ the Ho-moment quickly saturates and there is also a sharp spin-reorientation of Ho and Ru-moments, which gives rise to a maximum in the temperature dependent magnetic susceptibility and a large change in entropy leading to an anomaly in heat capacity. The absence of anomaly in magnetic susceptibility and heat capacity at the onset of magnetic ordering is unusual but not rare. Haldane spin-chain system ($R_2BaNiO_5$) exhibits similar features,[5] where long-range magnetic ordering develops at high temperature as revealed by neutron diffraction measurement, but it does not yield an anomaly in magnetic susceptibility and heat capacity until spin-reorientation and spin-saturation occur at lower temperature. Second, Ho and Ru spins simultaneously develop long range ordering below $T_{N1}$. In general, rare-earth ions often order at relatively low temperature because of the weak magnetic correlation due to their localized $f$ orbitals. For instance, $Ho_2Ru_2O_7$ undergoes two magnetic phase transitions with Ru moment ordered at higher temperature (~ 95 K) followed by the ordering of Ho ions at lower temperature (~ 1.4 K) due to the enhanced internal magnetic field arising from the ordered Ru sublattice.[8,9] The concurrent ordering of Ho and Ru moment at 50 K in $Ba_3HoRu_2O_9$ in the current study signals stronger Ho($4f$)-Ru($4d$) magnetic correlation. Based on Goodenough-Anderson-Kanamori rules, the dominant nearest-neighbor exchange interactions in this system include i) strong 179° Ru-O-Ho antiferromagnetic super-exchange interaction (see crystal structure in Fig. S6 in SM)[32], ii) - 78° Ru-O-Ho antiferromagnetic super-exchange interaction, and iii) weak Ru-Ru ferromagnetic direct exchange interaction ( Ru-Ru of a dimer ~2.55 Å which is less than Ru-Ru distance (2.65 Å) in a metal). The dominant 179° Ru-O-Ho antiferromagnetic super-exchange compared to Ru-O-Ru and Ru-Ru magnetic interaction in $Ba_3Ho^{+3}Ru^{+4.5}{}_2O_9$ could be one possible reason of simultaneous magnetic ordering of Ru and Ho-ions compared to that of the $Ho_2Ru_2O_7$ system (where Ho-O-Ru and Ru-O-Ru both exhibits ~ 109° super-exchange interaction).[9] The light $R$-member $Ba_3NdRu_2O_9$ in this family exhibits FM ordering below 24 K, followed by another magnetic ordering ~17-18 K. The Nd-moments align ferromagnetically below 24 K associated with a (0 0 0) wave vector and become canted antiferromagnetically ordered below 17 K with the same K-vector, whereas $Ru_2O_9$-dimers order antiferromagnetically with a (0.5 0 0) wave vector,[20] unlike this compound. The Ru-moments are canted in a $Ru_2O_9$ dimer in the titled system, unlike FM arrangement of Ru-moments in intra-dimer in $Ba_3NdRu_2O_9$ compound,[20] and unlike the earlier prediction of AFM-dimer in this family for all $R$-members.[17,28,29]

Third, two distinct magnetic phases coexist below $T_{N2}$, as illustrated in Fig. 3(b,c), one is associated with $\mathbf{K_1}$ = (0.5 0 0) and the other is associated with $\mathbf{K_2}$ = (0.25 0.25 0). Each $\mathbf{K}$ vector corresponds to magnetic ordering of both Ru and Ho simultaneously, unlike many other complex systems where two different $\mathbf{K}$-vectors are associated respectively with two different magnetic sublattices for different atoms or different crystallographic sites of the same atom.[34–36] Simultaneous ordering of both TM and $R$ ions at high temperature is also reported in the multiferroic compounds $RMn_2O_5$ [3,37–39] and $R_2BaNiO_5$[4,5], which exhibit successive magnetic anomalies at lower temperature. However, no such magnetic phase coexistence was observed at a particular same $T$-regime for these oxides. Magnetic phase coexistence below metal-insulator transition temperature was observed in a layered ruthenate system, Fe-doped $Ca_3Ru_2O_7$, where the coexistence of commensurate and incommensurate phases arises from competing FM and AFM in-plane Ru-Ru interactions [40]. Note that the incommensurate phase stems from the commensurate one as a result of modification in the FM exchange interaction due to Fe doping. In contrast, here we demonstrate an unconventional magnetic phase coexistence in a magnetically three-dimensional compound, $Ba_3HoRu_2O_9$, where two coexisting magnetic phases are completely different and the phase competition arises as a result of different competing exchange interactions. It is likely that the Ho-O-Ru super-exchange interaction starts to dominate below $T_{N2}$ and Ru-moments align with stronger Ho-moments, which results in spin-reorientation and the emergence of another spin configuration within the ab-plane by minimizing the exchange frustration. However, other parameters, such as crystal-field effect, magnetic anisotropy, may play a role as well, which needs further theoretical/spectroscopic investigations. An external parameter, like magnetic field or pressure, may stabilize to a particular magnetic phase by tuning the competing interactions.

The Lorentzian part in the peak-shape of magnetic reflections associated with $\mathbf{K_2}$ indicates a shorter magnetic correlation length compared to that of an ideal (perfect) 3D long-range-ordered magnet. However, λ-shape anomaly in heat-capacity below $T_{N2}$ is consistent with LRO as observed in a typical LRO system. It is likely that the system forms finite-size magnetic domains instead of a perfect LRO below $T_{N2}$ associated with $\mathbf{K_2}$. Such finite-size magnetic domains (having up-up-down-down spin-structure) instead of true LRO at AFM-ordering($T_N$) have been predicted for the well-known multiferroic compound $Ca_3CoMnO_6$, in which the small ferroelectric polarization (compared to that of theoretically calculated value) is considered due to cancellations of polarization originating from different magnetic domains.[33] The occurrence of small ferroelectric polarization for our titled compound below $T_{N2}$ (see Ref.[25]) may be justified using same rationale if the ferroelectricity stems from the up-up-down-down spin-structure.

In summary, we have revisited magnetic properties of multiferroic compound $Ba_3HoRu_2O_9$ via comprehensive neutron powder diffraction measurements. We find that this material undergoes two magnetic phase transitions, instead of one, at $T_{N1} \approx 50$ K and $T_{N2} \approx 10$ K, where both Ho and Ru spins develop long range order simultaneously. This suggests a strong $4d$ (Ru)-$4f$ (Ho) magnetic correlation. In addition, below $T_{N2}$ we unravel a coexistence of two magnetic phases associated with completely two different propagation wave vectors, implying competition among different exchange interactions. The exchange-striction from up-up-down-down structure could be the possible reason for observation of polarization below $T_{N2}$. This study demonstrates that $Ba_3RRu_2O_9$ system provides a unique platform to study the cooperative $4d$-$4f$ phenomena where Ru and $R$ moments are strongly coupled and compete with other exchange interactions. Our new interesting results further call for reinvestigation of magnetic orderings in other $R$-Ru based systems in general.

Work at Michigan State University was supported by the U.S. Department of Energy, Office of Science, Office of Basic Energy Sciences, Materials Sciences and Engineering Division under Award No. DE-SC0019259.

**Table-1:** The magnetic moment of Ru and Ho and its projection along different axis and angle.

| T (K) | K₁= (0.5 0 0) | | | | | | | | Angle with c-axis (⁰) | |
|---|---|---|---|---|---|---|---|---|---|---|
| | Ho-Moment (μ$_B$) | | | | Ru-Moment (μ$_B$) | | | | | |
| | M$_a$ | M$_b$ | M$_c$ | M$_{Ho}$ | M$_a$ | M$_b$ | M$_c$ | M$_{Ru}$ | M$_{Ho}$ | M$_{Ru}$ |
| 80 | - | - | - | - | - | - | - | - | | |
| 30 | 1.197 | 0.599 | -0.102 | 1.042 | -0.289 | -0.144 | 0.778 | 0.817 | 94.4 | 22.5 |
| 19 | 1.452 | 0.726 | -0.484 | 1.347 | -0.717 | -0.359 | 1.262 | 1.406 | 106.6 | 32.4 |
| 11.3 | 1.544 | 0.772 | -0.648 | 1.486 | -0.761 | -0.380 | 1.443 | 1.586 | 110.6 | 30.5 |
| 8.1 | 3.447 | 1.723 | 3.951 | 4.952 | -0.510 | -0.255 | -0.281 | 0.524 | 44.3 | 116.3 |
| 1.5 | 4.089 | 2.044 | 4.665 | 5.856 | -0.489 | -0.245 | -0.388 | 0.574 | 44.4 | 54.6 |
| K₂= (0.25 0.25 0)   T < T$_{N2}$ | | | | | | | | | | |
| 8.1 | 4.127 | 4.127 | 0 | 4.127 | 1.245 | 0.865 | 0 | 1.105 | 90 | 90 |
| 1.5 | 4.920 | 4.920 | 0 | 4.920 | 1.148 | 0.531 | 0 | 0.995 | 90 | 90 |

**Figure Captions:**

**Fig. 1:** (a) Heat Capacity as a function of temperature measured at $H = 0$ and 5 T. Inset of (a) shows the magnetic susceptibility as a function of temperature measured at $H = 1$ T. (b) Powder neutron diffraction pattern collected at different temperatures from 1.5-80 K in low Q-region. Selected magnetic Bragg peaks are indexed, as described in the text, and the insets show the expanded views. (c) Magnetic Bragg peak intensity of (0.5 0 2) plotted as function of temperature measured at zero field. The inset shows the ordering parameter measurements of (0.75 -0.25 1), (0.5 0 1) and (0.25 0.25 1) magnetic Bragg peaks.

**Fig. 2:** Powder neutron diffraction pattern collected at $T$= 80 K (a), 30 K (b), 11.3 K (c) and 8.1 K (d) in zero magnetic field. The open black circle represents the experimental data, while the red solid line shows the Rietveld fitting. The vertical bars display the Bragg peak positions. The upper vertical lines represent Bragg peaks of crystal structure of Ba$_3$HoRu$_2$O$_9$, the next lower vertical lines represent magnetic Bragg peaks associated with **K**$_1$ = (0.5 0 0) (for (b), (c), and (d)); and lowest vertical line in (d) represents magnetic Bragg peaks associated with **K**$_2$ = (0.25 0.25 0). The continuous dark yellow at the bottom of the figure shows the difference between the experimental and calculated intensity.

**Fig. 3:** (a) and (b) represent magnetic structure at $T = 30$ K and 11.3 K, respectively. (c) and (d) represent magnetic structure at 8.1 K, associated with **K**$_1$ and **K**$_2$, respectively. The length of magnetic vectors represents the relative moment size of Ho (blue) and Ru (green) at that particular temperature.

Fig. 1.

T. Basu et al,

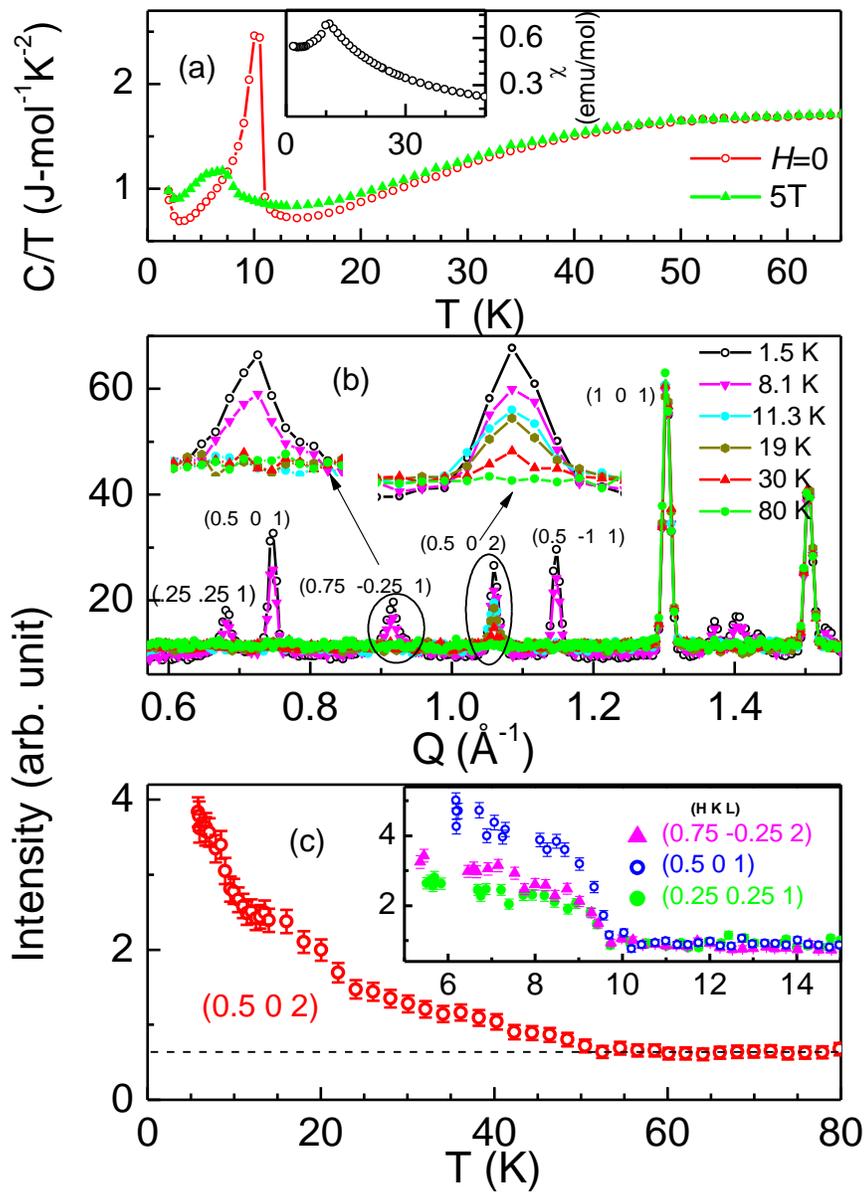

Fig. 2.

T. Basu et al,

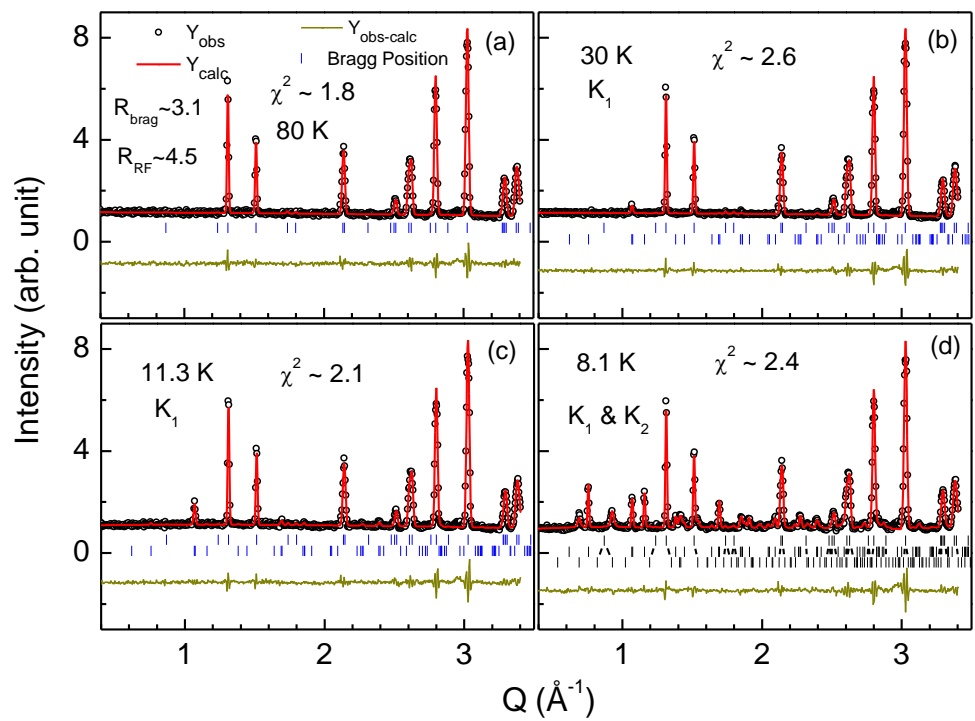

Fig. 3.

T. Basu et al,

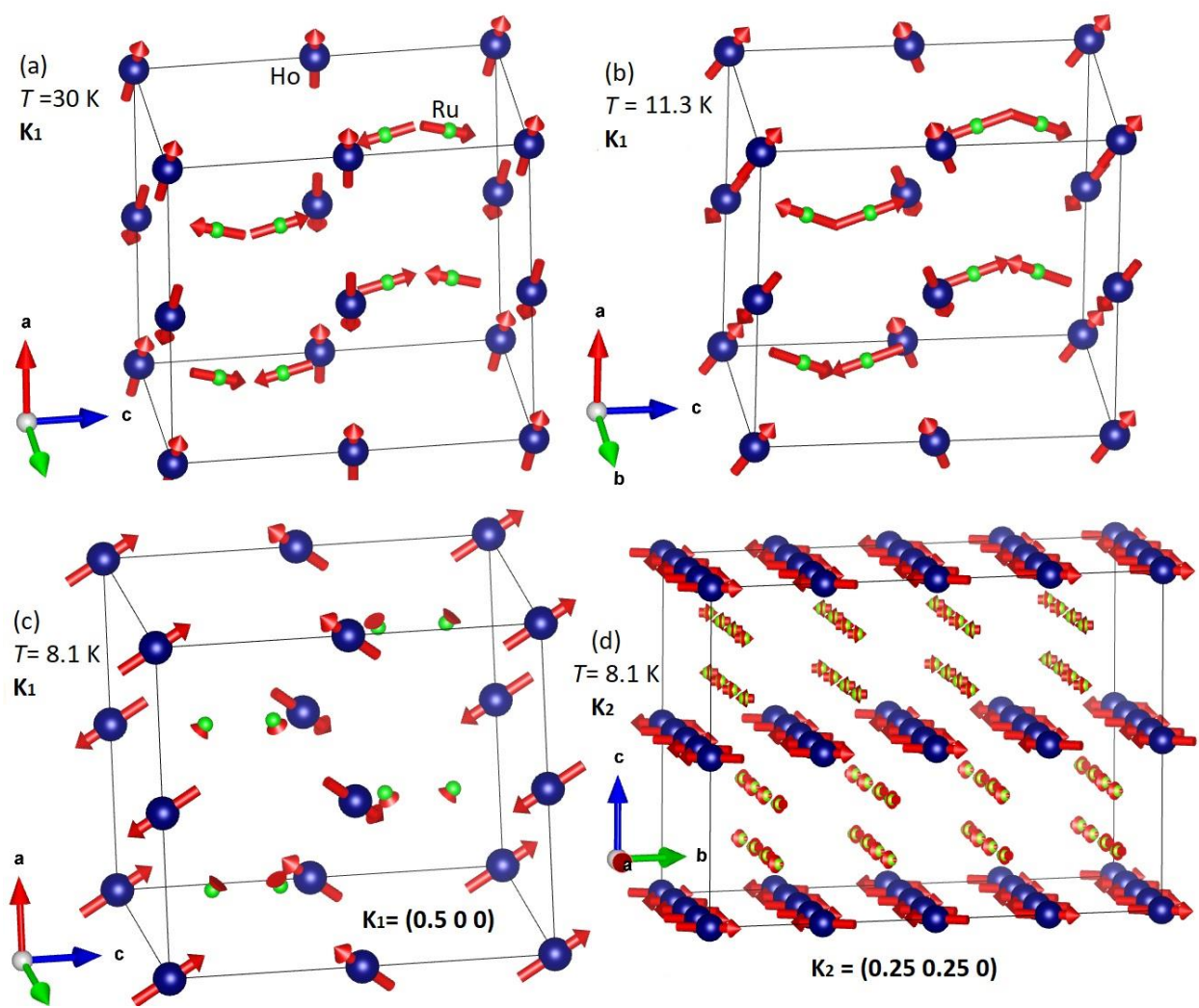

# Supplemental Material

## A. *Magnetic susceptibility and Heat Capacity:*

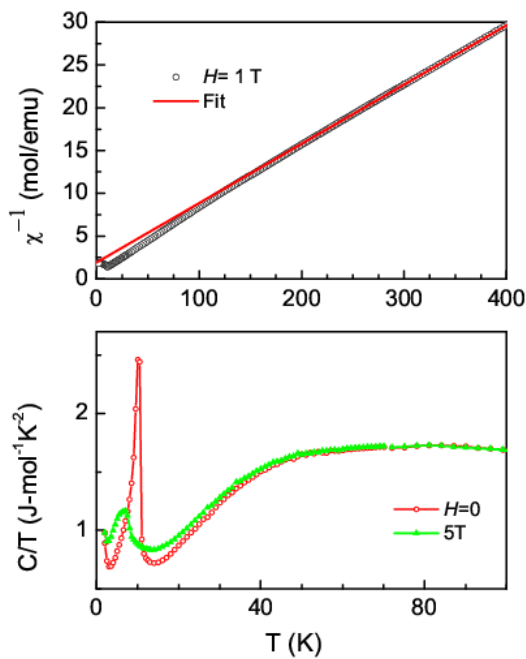

**Fig. S1:** (a) Inverse magnetic susceptibility as a function of temperature in presence of 1 T magnetic field from 2-320 K. The solid red line is a linear fit. (b) Heat Capacity divided by temperature as a function of temperature for zero magnetic field.

The obtained effective moment ($\mu_{eff}$) and Weiss ($\Theta_P$) temperature, from Curie-Weiss fit between 250-400 K, is 10.8 $\mu_B$ and -28 K, respectively. Considering the full moment of Ho$^{+3}$ (10.6 $\mu_B$), the effective moment of Ru (1.4 $\mu_B$) is significantly reduced compare to its theoretical value (2.4 $\mu_B$) of spin-only moment (S=1 for Ru$^{+4}$ and S=3/2 for Ru$^{+5}$). However, short-range magnetic correlations may exist even at very high temperature in this dimmer system, thereby, a true paramagnetic region could not be obtained which could include errors in fitting results.

## B. *Neutron and Magnetic Structural Analysis using SARAh and Fullprof program:*

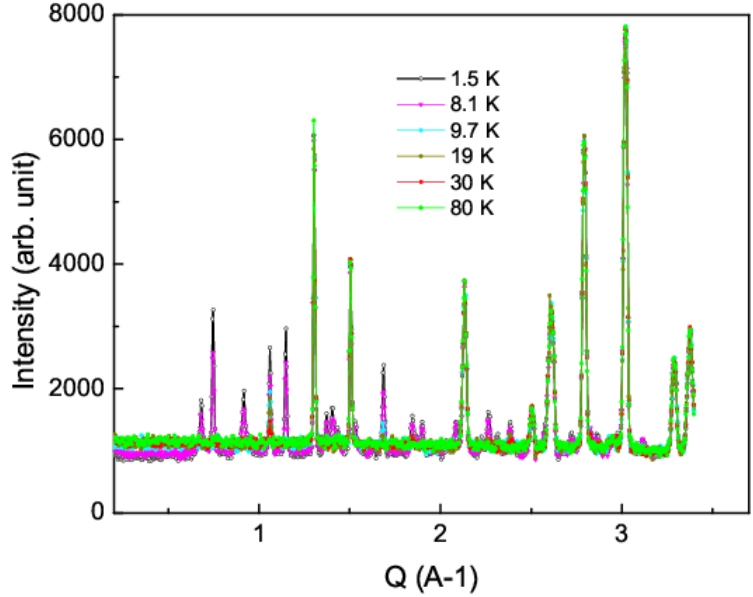

**Fig. S2:** Neutron diffraction intensity at several fermetures for large Q-range to get a better view of no change of magnetic intensity at nuclear Bragg peaks.

## Magnetic moments on both Ru and Ho-moments: Detail of Refinement:

The magnetic reflection associated with $\mathbf{K_1} = (0.5\ 0\ 0)$ is best fitted only if we consider magnetic moments on both Ru and Ho-moments. However, we have tried different possibilities, i) one with fixing magnetic moment of Ho equal to zero and refining the magnetic moment with Ru-atom only (Case-I), and ii) another with fixing the magnetic moment of Ru what is obtained in Case-I (Or, restricted to theoretical value) and then refine the magnetic moments of Ho, iii) the 3$^{rd}$ one is refining both the magnetic moments on Ru and Ho-atom. The three cases of the refinement are depicted in Fig.S3 for $T = 11.3$ K. The case-I (Fig. S3-a) gives the worse fitting of the magnetic reflection of (0.5 0 1) compared to that of case-II (Fig. S3-b) and case-III (Fig. S3-a). The magnetic moment of Ru obtained from case-I is ~2.4 $\mu_B$ which is equal to the theoretical saturation moment considering spin-only value but much higher than experimentally obtained Ru-moment from magnetic susceptibility. Case-III is slightly better fitted compared to that of case-II. We observe the similar behavior for all other temperatures for $T_{N2} < T < T_{N1}$ (not shown here). Therefore, these results confirm the simultaneous ordering of Ru and Ho-moments. Further, we elaborate all these possibilities below $T_{N2}$. One argument could be that the Ru-moments order at higher temperature below $T_{N1}$, followed by another magnetic structure associated with Ho-moments. If we consider the first case, then Ru-moment at 11.3 K is already saturated (as obtained from refinement in case-I and case-II), therefore, we fix the theoretical value of Ru-moments (2.4 $\mu_B$) and refine the magnetic structure by varying Ho-moment at 8.1 K (Fig. S3-d). Not only the magnetic reflection (0.5 0 1) is badly fitted, but another magnetic reflection, which is associated with propagation vector **K₁**, (0.5 -1 1), gives negligible intensity in this modeling compared to that of experimentally obtained value. Whereas, the latter case, that is, refinement on both Ho and Ru-moments, gives nearly exact fitting to all the magnetic reflections associated with **K₁** (Fig. S3-e). Thereby, as discussed in the main manuscript, the neutron data at $T_{N2} < T < T_{N1}$ are best modeled with magnetic ordering of both Ru and Ho. The obtained Ho and Ru magnetic moments are tabulated in Table-I in main manuscript.

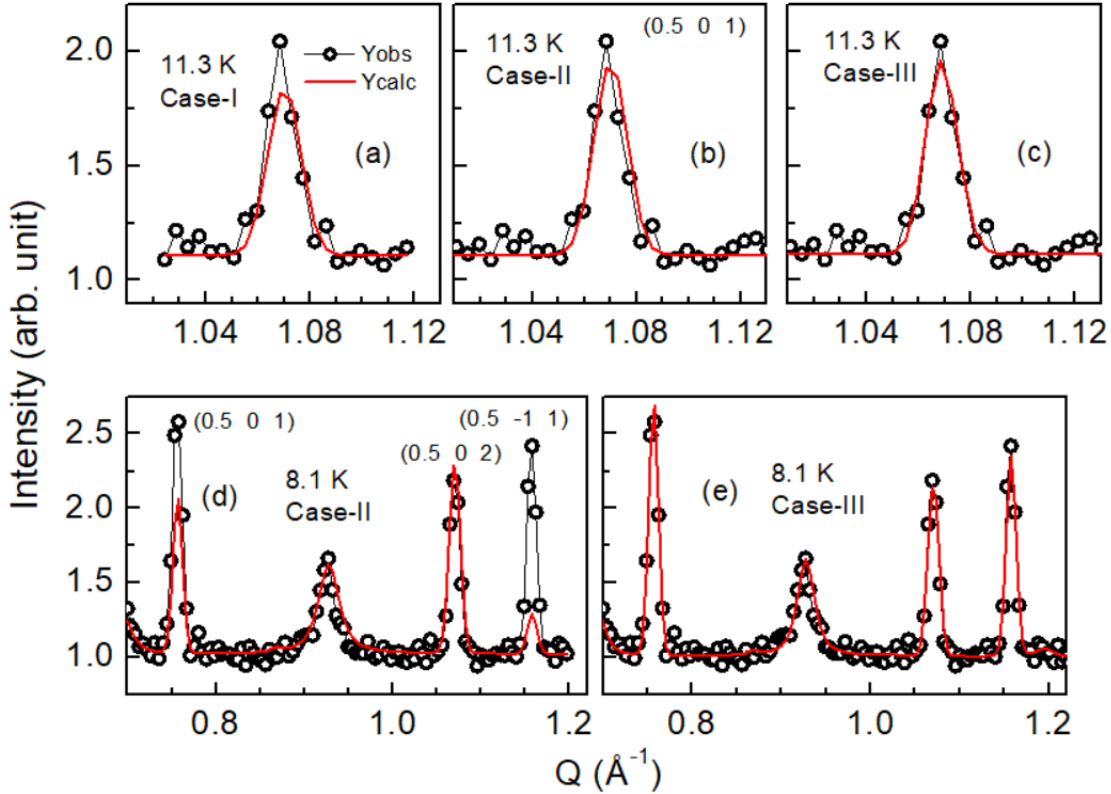

**Figures S3:** Rietveld refinement at 11.3 K (a, b,c) and 8.1 K (d,e), as discussed in the text. Case-I: Refine only Ru-moment, keeping zero moment on Ho. Case-II: Fixing Ru-moment to its theoretical value/ what is obtained in case-I and refine the moment on Ho. Case-III: Refine moments on both Ru and Ho simultaneously.

**Rietveld Refinement and magnetic structure at T=1.5 K:**

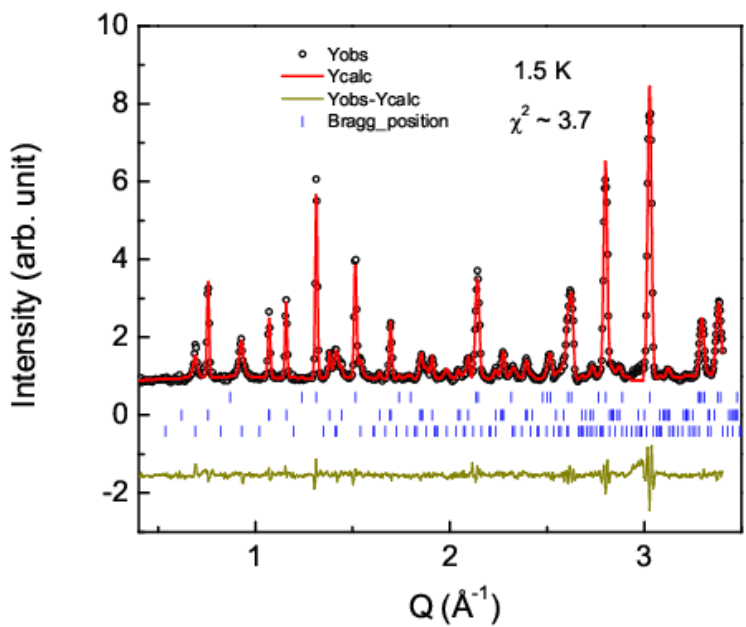

**Fig. S4:** Rietveld refinement at 1.5 K

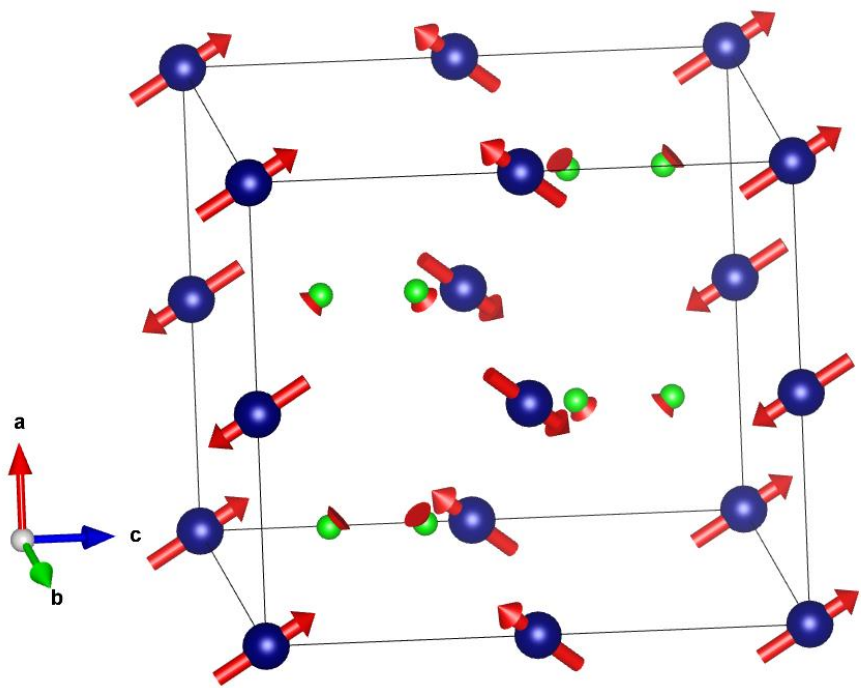

**Fig. S5:** Magnetic structure at 1.5 K associated with $K_1$= 0.5 0 0.

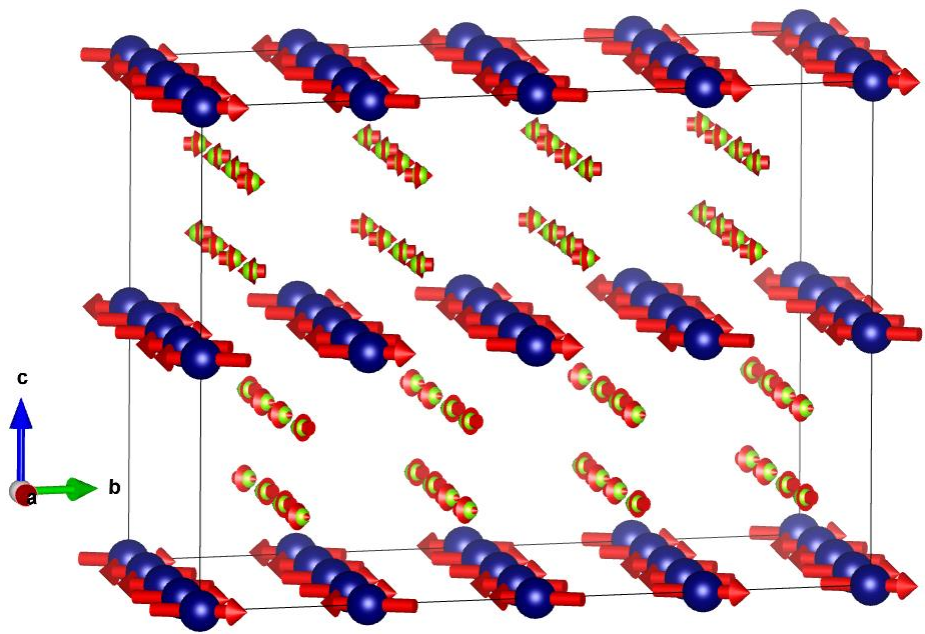

**Fig. S6:** Magnetic structure at 1.5 K associated with **K₂**=0.5 0.5 0

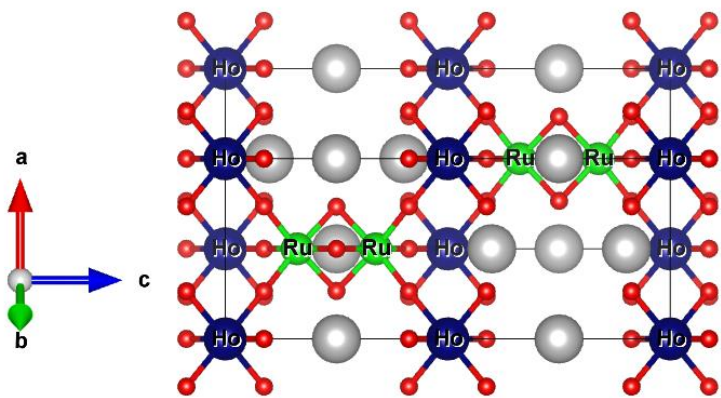

**Figures S6: Crystal structure of Ba$_3$HoRu$_2$O$_9$.**

**Table-S1: Allowed k-vector for space group P63/mmc.**

| k-vector description | | ITA description | |
|---|---|---|---|
| k-vector label | Conventional basis | Wyckoff position | |
| | | Multiplicity | Letter |
| GM | 0,0,0 | 1 | a |
| A | 0,0,1/2 | 1 | b |
| K | 1/3,1/3,0 | 2 | c |
| H | 1/3,1/3,1/2 | 2 | d |
| DT | 0,0,u | 2 | e |
| M | 1/2,0,0 | 3 | f |
| L | 1/2,0,1/2 | 3 | g |
| P | 1/3,1/3,u | 4 | h |
| U | 1/2,0,u | 6 | i |
| SM | u,0,0 | 6 | j |
| R | u,0,1/2 | 6 | k |
| LD (T) | u,u,0 | 6 | l |
| Q (S) | u,u,1/2 | 6 | m |
| D | u,0,v | 12 | n |
| C (F) | u,u,v | 12 | o |
| B | u,v,0 | 12 | p |
| E | u,v,1/2 | 12 | q |
| GP | u,v,w | 24 | r |

**Table-S2:** Basis vector for the space group P63/mmc for K=0.5 0 0. The irreducible representation for magnetic site Ru (0.66667 0.33333 0.16164), as obtained from SARAh program.

| IR | BV | Atom | BV components | | | | | |
|---|---|---|---|---|---|---|---|---|
| | | | $m_{\|a}$ | $m_{\|b}$ | $m_{\|c}$ | $im_{\|a}$ | $im_{\|b}$ | $im_{\|c}$ |
| $\Gamma_1$ | $\psi_1$ | 1 | 0 | -1 | 0 | 0 | 0 | 0 |
| | | 2 | 0 | -1 | 0 | 0 | 0 | 0 |
| | | 3 | 0 | 1 | 0 | 0 | 0 | 0 |
| | | 4 | 0 | 1 | 0 | 0 | 0 | 0 |
| $\Gamma_2$ | $\psi_2$ | 1 | 2 | 1 | 0 | 0 | 0 | 0 |
| | | 2 | 2 | 1 | 0 | 0 | 0 | 0 |
| | | 3 | 2 | 1 | 0 | 0 | 0 | 0 |
| | | 4 | 2 | 1 | 0 | 0 | 0 | 0 |
| | $\psi_3$ | 1 | 0 | 0 | 2 | 0 | 0 | 0 |
| | | 2 | 0 | 0 | -2 | 0 | 0 | 0 |
| | | 3 | 0 | 0 | 2 | 0 | 0 | 0 |
| | | 4 | 0 | 0 | -2 | 0 | 0 | 0 |
| $\Gamma_3$ | $\psi_4$ | 1 | 2 | 1 | 0 | 0 | 0 | 0 |
| | | 2 | 2 | 1 | 0 | 0 | 0 | 0 |
| | | 3 | -2 | -1 | 0 | 0 | 0 | 0 |
| | | 4 | -2 | -1 | 0 | 0 | 0 | 0 |
| | $\psi_5$ | 1 | 0 | 0 | 2 | 0 | 0 | 0 |
| | | 2 | 0 | 0 | -2 | 0 | 0 | 0 |
| | | 3 | 0 | 0 | -2 | 0 | 0 | 0 |
| | | 4 | 0 | 0 | 2 | 0 | 0 | 0 |
| $\Gamma_4$ | $\psi_6$ | 1 | 0 | -1 | 0 | 0 | 0 | 0 |
| | | 2 | 0 | -1 | 0 | 0 | 0 | 0 |
| | | 3 | 0 | -1 | 0 | 0 | 0 | 0 |
| | | 4 | 0 | -1 | 0 | 0 | 0 | 0 |
| $\Gamma_5$ | $\psi_7$ | 1 | 0 | -1 | 0 | 0 | 0 | 0 |
| | | 2 | 0 | 1 | 0 | 0 | 0 | 0 |
| | | 3 | 0 | 1 | 0 | 0 | 0 | 0 |
| | | 4 | 0 | -1 | 0 | 0 | 0 | 0 |
| $\Gamma_6$ | $\psi_8$ | 1 | 2 | 1 | 0 | 0 | 0 | 0 |
| | | 2 | -2 | -1 | 0 | 0 | 0 | 0 |
| | | 3 | 2 | 1 | 0 | 0 | 0 | 0 |
| | | 4 | -2 | -1 | 0 | 0 | 0 | 0 |
| | $\psi_9$ | 1 | 0 | 0 | 2 | 0 | 0 | 0 |
| | | 2 | 0 | 0 | 2 | 0 | 0 | 0 |
| | | 3 | 0 | 0 | 2 | 0 | 0 | 0 |
| | | 4 | 0 | 0 | 2 | 0 | 0 | 0 |
| $\Gamma_7$ | $\psi_{10}$ | 1 | 2 | 1 | 0 | 0 | 0 | 0 |
| | | 2 | -2 | -1 | 0 | 0 | 0 | 0 |
| | | 3 | -2 | -1 | 0 | 0 | 0 | 0 |
| | | 4 | 2 | 1 | 0 | 0 | 0 | 0 |

| IR | BV | Atom | BV components | | | | | |
|---|---|---|---|---|---|---|---|---|
| | | | $m_a$ | $m_b$ | $m_c$ | $im_a$ | $im_b$ | $im_c$ |
| | $\psi_{11}$ | 1 | 0 | 0 | 2 | 0 | 0 | 0 |
| | | 2 | 0 | 0 | 2 | 0 | 0 | 0 |
| | | 3 | 0 | 0 | -2 | 0 | 0 | 0 |
| | | 4 | 0 | 0 | -2 | 0 | 0 | 0 |
| $\Gamma_8$ | $\psi_{12}$ | 1 | 0 | -1 | 0 | 0 | 0 | 0 |
| | | 2 | 0 | 1 | 0 | 0 | 0 | 0 |
| | | 3 | 0 | -1 | 0 | 0 | 0 | 0 |
| | | 4 | 0 | 1 | 0 | 0 | 0 | 0 |

**Table-S3:** Basis vector for the space group P63/mmc for K=0.5 0 0. The irreducible representation for magnetic site Ho (0 0 0), as obtained from SARAh program.

| IR | BV | Atom | BV components | | | | | |
|---|---|---|---|---|---|---|---|---|
| | | | $m_{\parallel a}$ | $m_{\parallel b}$ | $m_{\parallel c}$ | $im_{\parallel a}$ | $im_{\parallel b}$ | $im_{\parallel c}$ |
| $\Gamma_1$ | $\psi_1$ | 1 | 0 | -2 | 0 | 0 | 0 | 0 |
| | | 2 | 0 | 2 | 0 | 0 | 0 | 0 |
| $\Gamma_3$ | $\psi_2$ | 1 | 4 | 2 | 0 | 0 | 0 | 0 |
| | | 2 | -4 | -2 | 0 | 0 | 0 | 0 |
| | $\psi_3$ | 1 | 0 | 0 | 4 | 0 | 0 | 0 |
| | | 2 | 0 | 0 | 4 | 0 | 0 | 0 |
| $\Gamma_5$ | $\psi_4$ | 1 | 0 | -2 | 0 | 0 | 0 | 0 |
| | | 2 | 0 | -2 | 0 | 0 | 0 | 0 |
| $\Gamma_7$ | $\psi_5$ | 1 | 4 | 2 | 0 | 0 | 0 | 0 |
| | | 2 | 4 | 2 | 0 | 0 | 0 | 0 |
| | $\psi_6$ | 1 | 0 | 0 | 4 | 0 | 0 | 0 |
| | | 2 | 0 | 0 | -4 | 0 | 0 | 0 |

**Table-S4: Basis vector for the space group P63/mmc for K=0.25 0.25 0. The irreducible representation for magnetic site Ru (0.66667 0.33333 0.16164), as obtained from SARAh program.**

| IR | BV | Atom | BV components | | | | | |
|---|---|---|---|---|---|---|---|---|
| | | | $m_{\parallel a}$ | $m_{\parallel b}$ | $m_{\parallel c}$ | $im_{\parallel a}$ | $im_{\parallel b}$ | $im_{\parallel c}$ |
| $\Gamma_1$ | $\psi_1$ | 1 | 1 | 0 | 0 | 0 | 0 | 0 |
| | | 2 | 0 | 1 | 0 | 0 | 0 | 0 |
| | | 3 | -1 | 0 | 0 | 0 | 0 | 0 |
| | | 4 | 0 | -1 | 0 | 0 | 0 | 0 |
| | $\psi_2$ | 1 | 0 | 1 | 0 | 0 | 0 | 0 |
| | | 2 | 1 | 0 | 0 | 0 | 0 | 0 |
| | | 3 | 0 | -1 | 0 | 0 | 0 | 0 |
| | | 4 | -1 | 0 | 0 | 0 | 0 | 0 |
| | $\psi_3$ | 1 | 0 | 0 | 1 | 0 | 0 | 0 |
| | | 2 | 0 | 0 | -1 | 0 | 0 | 0 |
| | | 3 | 0 | 0 | 1 | 0 | 0 | 0 |
| | | 4 | 0 | 0 | -1 | 0 | 0 | 0 |
| $\Gamma_2$ | $\psi_4$ | 1 | 1 | 0 | 0 | 0 | 0 | 0 |
| | | 2 | 0 | 1 | 0 | 0 | 0 | 0 |
| | | 3 | 1 | 0 | 0 | 0 | 0 | 0 |
| | | 4 | 0 | 1 | 0 | 0 | 0 | 0 |
| | $\psi_5$ | 1 | 0 | 1 | 0 | 0 | 0 | 0 |
| | | 2 | 1 | 0 | 0 | 0 | 0 | 0 |
| | | 3 | 0 | 1 | 0 | 0 | 0 | 0 |
| | | 4 | 1 | 0 | 0 | 0 | 0 | 0 |
| | $\psi_6$ | 1 | 0 | 0 | 1 | 0 | 0 | 0 |
| | | 2 | 0 | 0 | -1 | 0 | 0 | 0 |
| | | 3 | 0 | 0 | -1 | 0 | 0 | 0 |
| | | 4 | 0 | 0 | 1 | 0 | 0 | 0 |
| $\Gamma_3$ | $\psi_7$ | 1 | 1 | 0 | 0 | 0 | 0 | 0 |
| | | 2 | 0 | -1 | 0 | 0 | 0 | 0 |
| | | 3 | -1 | 0 | 0 | 0 | 0 | 0 |
| | | 4 | 0 | 1 | 0 | 0 | 0 | 0 |
| | $\psi_8$ | 1 | 0 | 1 | 0 | 0 | 0 | 0 |
| | | 2 | -1 | 0 | 0 | 0 | 0 | 0 |
| | | 3 | 0 | -1 | 0 | 0 | 0 | 0 |
| | | 4 | 1 | 0 | 0 | 0 | 0 | 0 |
| | $\psi_9$ | 1 | 0 | 0 | 1 | 0 | 0 | 0 |
| | | 2 | 0 | 0 | 1 | 0 | 0 | 0 |
| | | 3 | 0 | 0 | 1 | 0 | 0 | 0 |
| | | 4 | 0 | 0 | 1 | 0 | 0 | 0 |
| $\Gamma_4$ | $\psi_{10}$ | 1 | 1 | 0 | 0 | 0 | 0 | 0 |
| | | 2 | 0 | -1 | 0 | 0 | 0 | 0 |
| | | 3 | 1 | 0 | 0 | 0 | 0 | 0 |
| | | 4 | 0 | -1 | 0 | 0 | 0 | 0 |

| IR | BV | Atom | BV components ||||||
|---|---|---|---|---|---|---|---|---|
| | | | $m_a$ | $m_b$ | $m_c$ | $im_a$ | $im_b$ | $im_c$ |
| | $\psi_{11}$ | 1 | 0 | 1 | 0 | 0 | 0 | 0 |
| | | 2 | -1 | 0 | 0 | 0 | 0 | 0 |
| | | 3 | 0 | 1 | 0 | 0 | 0 | 0 |
| | | 4 | -1 | 0 | 0 | 0 | 0 | 0 |
| | $\psi_{12}$ | 1 | 0 | 0 | 1 | 0 | 0 | 0 |
| | | 2 | 0 | 0 | 1 | 0 | 0 | 0 |
| | | 3 | 0 | 0 | -1 | 0 | 0 | 0 |
| | | 4 | 0 | 0 | -1 | 0 | 0 | 0 |

**Table-S5: Basis vector for the space group P63/mmc for K=0.25 0.25 0. The irreducible representation for magnetic site Ho (0 0 0), as obtained from SARAh program.**

| IR | BV | Atom | BV components ||||||
|---|---|---|---|---|---|---|---|---|
| | | | $m_{\|a}$ | $m_{\|b}$ | $m_{\|c}$ | $im_{\|a}$ | $im_{\|b}$ | $im_{\|c}$ |
| $\Gamma_1$ | $\psi_1$ | 1 | 1 | 1 | 0 | 0 | 0 | 0 |
| | | 2 | -1 | -1 | 0 | 0 | 0 | 0 |
| $\Gamma_2$ | $\psi_2$ | 1 | 1 | 1 | 0 | 0 | 0 | 0 |
| | | 2 | 1 | 1 | 0 | 0 | 0 | 0 |
| $\Gamma_3$ | $\psi_3$ | 1 | 1 | -1 | 0 | 0 | 0 | 0 |
| | | 2 | -1 | 1 | 0 | 0 | 0 | 0 |
| | $\psi_4$ | 1 | 0 | 0 | 2 | 0 | 0 | 0 |
| | | 2 | 0 | 0 | 2 | 0 | 0 | 0 |
| $\Gamma_4$ | $\psi_5$ | 1 | 1 | -1 | 0 | 0 | 0 | 0 |
| | | 2 | 1 | -1 | 0 | 0 | 0 | 0 |
| | $\psi_6$ | 1 | 0 | 0 | 2 | 0 | 0 | 0 |
| | | 2 | 0 | 0 | -2 | 0 | 0 | 0 |

**Table-S6:** Calculated and experimentally obtained intensity, obtained from riveted refinement in Fullprof Suite program, for particular peaks for different temperature.

| H K L | Associated k-vector | T (K) | $I_{calc}$ | $I_{expt}$ |
|---|---|---|---|---|
| 0.5 0 1 | ½ 0 0 | 1.5 | 374.6 | 383.8 |
| | | 8.1 | 255.9 | 260.9 |
| | | 11.3 | 11.4 | 5.1 |
| | | 19 | 5.6 | 3.6 |
| | | 30 | 1.4 | 0.0 |
| 0.5 0 2 | | 1.5 | 121.6 | 131.1 |
| | | 8.1 | 90.6 | 93.8 |
| | | 11.3 | 96.0 | 114.8 |
| | | 19 | 80.0 | 84.5 |
| | | 30 | 38.5 | 40.5 |
| 0.5 -1 1 | | 1.5 | 306.3 | 328.7 |
| | | 8.1 | 210.7 | 217.8 |
| | | 11.3 | 5.0 | 7.9 |
| | | 19 | 5.0 | 1.1 |
| | | 30 | 4.2 | 3.5 |
| 0.25 0.25 1 | ¼ ¼ 0 | 1.5 | 220.4 | 206.1 |
| | | 8.1 | 141.3 | 132.2 |
| 0.75 -0.25 1 | | 1.5 | 185.8 | 181.5 |
| | | 8.1 | 114.8 | 113.1 |